\documentclass[a4paper]{article}

\usepackage{INTERSPEECH2019}

\title{A neural network based post-filter for speech-driven head motion synthesis}
\name{JinHong Lu, Hiroshi Shimodaira}
\address{Centre for Speech Technology Research, School of Informatics,\\ University of Edinburgh, United Kingdom}
\email{J.Lu-20@sms.ed.ac.uk, H.Shimodaira@ed.ac.uk}

\begin{document}

\maketitle
\begin{abstract}
Despite the fact that neural networks are widely used for speech-driven head motion synthesis, it is well-known that the output of neural networks is noisy or discontinuous due to the limited capability of deep neural networks in predicting human motion. Thus, post-processing is required to obtain smooth head motion trajectories for animation. It is common to apply a linear filter or consider keyframes as post-processing. However, neither approach is optimal as there is always a trade-off between smoothness and accuracy. We propose to employ a neural network trained in a way that it is capable of reconstructing the head motions, in order to overcome this limitation. In the objective evaluation, this filter is proved to be good at de-noising data involving types of noise (dropout/Gaussian noise). Objective metrics also demonstrate improvement of the joined head motion’s smoothness after being processed by our proposed filter. A detailed analysis reveals that our proposed filter learns the characteristic of head motions. The subjective evaluation shows that participants were unable to distinguish the synthesised head motions with our proposed filter from ground truth, which was preferred over the Gaussian filter and moving average.
\end{abstract}
\noindent\textbf{Index Terms}: neural networks, post-filter, head motion synthesis, talking avatar

\section{Introduction}
\label{sec:intro}
Predicting human motions using deep neural networks has snowballed and achieved high success in terms of synchronicity between the ground truth and predicted one. Some examples include hand gesture and eye-gaze recognition and generation ~\cite{Stefanov2018RecognitionAG}, human motion for long-term prediction~\cite{GhoshSAH17}, and speech-driven head motion prediction~\cite{Haag2016}~\cite{Ding2015}. Human motion is continuous and regular, but most systems predict motion in short segments due to the learning capability of deep neural networks; there are many challenges in predicting motion from raw data over short time horizons – let alone long time horizons. This shortcoming results in the predicted movements being discontinuous and laggy or jerky. Hence, it is of paramount importance to have a de-noising/smoothing filter for these movements. 

There are two popular types of filters used for head motion synthesis research: linear filter (e.g. Gaussian filter, moving averaging smoothing) or de-noising auto-encoder using a neural network~\cite{GhoshSAH17}. The linear filters is a filter whose impulse response (or response to any finite length input) is of finite duration, because it settles to zero in finite time. De-noising auto-encoder is trained by inputting noisy data and computing a loss on the output by comparing it to the ground truth data. The disadvantage of the linear filters is that it does not have the additional information/knowledge of the characteristic of the actual movement track. The linear filters only use delayed versions of the input signal to filter the input to the output, and this may result in filtering the pivotal motion over the period, whereas the de-noising auto-encoder is trained with specific human movement data, which creates uniqueness and knowledge of the characteristic of the movement to the model. Thus, the keyframes of the movement have remained, while the noise is being removed.

In this paper, we propose an approach to de-noise head motion trajectories over a time period using an auto-encoder to recover the characteristic of the head movement track. Head motion is small and can move with either high ffrequency (e.g. repeated nodding/shaking) or low frequency (e.g. stillness); this creates ambiguities in the linear filters, which cannot identify whether the motion is the key frame or noise. Moreover, the output of NN-based regression speech-driven head motion synthesis is always very noisy due to the nature of speech; the NN-based system is unable to adapt and learn. It is common to apply post-processing to obtain a smooth output. Ding~\cite{Ding2015BLSTMNN} has applied MLPG~\cite{Tokuda2000} to generate smooth trajectories; Sadoughi~\cite{Sadoughi18} smoothened the rotations by converting into quaternions and then selecting 15 key points per second, interpolating the intermediate frames~\cite{Busso2007rigid}; and Hagg~\cite{Haag2016} applies 3-order polynomial smoothing filter on the output. However, these smoothing methods have the common problem that there is a trade-off between the smoothness of the filtered head motion and how accurate is the filtered head motion compare to the ground truth. The term ‘accurate’ here means that these post-filters may over-smooth the head motion and cause the filtered head motion is stirless. We propose a neural network based post-filter to overcome these problems by learning the characteristic of the head motion to reconstruct them.

The linear filters are based on identifying the impulse transfer function that satisfies the requirements of the filter specification, whereas our proposed filter requires inputting noisy data to train and learn in reconstructing smooth head motion. To the best of our knowledge, the common way to create noisy data is either applying dropout to the ‘clean’ data for making the data discontinuous or add Gaussian noise to the ‘clean’ data. These two methods are not suitable for creating noisy head motion data. Because the head motion only consists of three trajectories (X, Y, Z) in rotation vector, the dropout method drops one of the three trajectories of the head motion, and this strictly limits the movements causing unnatural behaviour and loss of information in the motion. On the other hand, adding Gaussian noise to create noisy data does not yield expected jerky movements as they would naturally occur. Thus, training an auto-encoder with Gaussian noise data would not be effective. Lastly, manually creating natural, jerky head motion is extremely expensive. Therefore, we would like to investigate how effective the de-noising auto-encoder can be by recreating the track of the ‘clean’ head motion. A distinct head motion can last over 400ms~\cite{Hofer2007AutomaticHM}, and the model has to ‘see’ multiple frames information at each time step to smoothen a complete head motion. Moreover, it is interesting to see how robust the model can be against the pure Gaussian noise or dropout noise added to the clean data. Lastly, how much improvement the de-noising auto-encoder makes as compared to the linear filters is another key exploration.

The remainder of the paper is organised as follows. Section 2 sets the proposed model for the prediction and filtration task. In section 3, we detail our dataset, implementation, and experiments. Section 4 discusses the objective evaluation based on the result in section 3. Subjective evaluation is discussed further in section 5. Lastly, we have a conclusion in section 6.

\begin{figure}[tb]
\centering
\centerline{\includegraphics[width=\columnwidth, height=4cm]{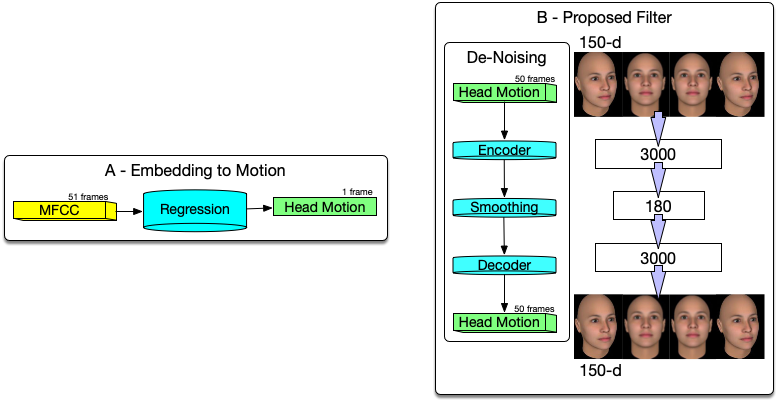}}
\caption{The overall framework for our proposed model. A:The regression model predicts head motion frame by frame from the stacking 51 frames of MFCC. B:The auto-encoder for de-noising the distinct head motion over 500ms. '-d': dimension of the data.}
\label{fig:proposed_model}
\vskip -5mm
\end{figure}

\section{Proposed Model}

Our proposed model can be represented as a de-noising auto-encoder for smoothing the predicted discontinuous head motions. The objective criterion for the model is the mean square error normalised by the variance of the ground truth for the training. The overall framework of our proposed model is illustrated in Figure~\ref{fig:proposed_model}(B).

\subsection{Auto-Encoder - De-noising Filter}

The common training procedure of the de-noising model, which comprises applying dropout/Gaussian noise to the clean data for recreating noisy data~\cite{GhoshSAH17}~\cite{Vincent:2010:SDA:1756006.1953039}, does not work with our model as the Gaussian noise method does not give the expected jerky movements as they would naturally occur. The dropout method, on the other hand, drops one of the three trajectories of the head motion, and this strictly limits the movement, causing unnatural behaviour. Therefore, instead of removing the noise from the jerky head motion, we expect the de-noising filter to learn and know how the smooth head motion over a period should be. We assume a complete head motion in every consecutive 500ms~\cite{Hofer2007AutomaticHM} time frame, as the input, $M_f$, to the de-noising filter and the output are of the same length. We follow the architecture in~\cite{GhoshSAH17}, using the feed-forward neural network, trained with the back-propagation learning algorithm, but as the input dimension is different in these two cases, we explored the best depth and width of the model for recovering the head motion. Overall, the filter can be represented by the following architecture:
\begin{equation}
M_f = W_{dl}(W_{el}M_f + b_{el})+ b_{dl} \quad \text{for} \quad 1 \leq l < L
\end{equation}
where $e$ represents the encoder operator, $d$ sets the decoder operator, and $l$ is the number of layers.

\section{Experiment}

\subsection{Data}
We use the University of Edinburgh Speaker Personality and Mocap Dataset~\cite{Haag2015TheUO}. This database contains expressive dialogues between semi-professional actors in extroverted and introverted speaking styles, and the dialogues were non-scripted and spontaneous. There is a total of 13 speakers, with 123 files in the data set, and each file is about 5 minutes long.

\textbf{Head Motion Features} The head motion of one speaker of the dialogue pair was recorded with the NaturalPoint Optitrack~\cite{NaturealPoint} motion capture system at a 100Hz sampling rate. From the marker coordinates, rotation matrices for head motion were computed using singular value decomposition~\cite{Soderkvist1994}. The rotation matrices were converted to rotation vectors, which describe the motions of X, Y, and Z. The head motion features are normalised for each dimension by mean and variant.

Furthermore, we assume there is a complete head motion in every consecutive 500ms and 250ms, which keeps shifting to ensure smoothness and continuity in every distinct head motion.

Moreover, we created two noisy datasets to examine the performance of the filter model: 1) Dropping (rate=0.5) frames of each 50-frame joint distinct head motion. These dropped frames have the values of zero in three trajectories. 2) Gaussian noise (0, 0.2) is added to the ‘clean’ dataset.

\subsection{Experimental Setups}
We conducted preliminary experiments to decide the architecture of the regression model for the prediction model in Figure~\ref{fig:proposed_model}(A). The prediction model used input and output features from the single speaker, who has 10 files in the data set. The speech data is extracted in the same way as mentioned in Jin~\cite{jin2019} and the corresponding six files are assigned for the training, two files for the validation and the remaining two files for the test data set. On the other hand, we use all 13 speakers’ head motion data to train the de-noising model because we assume that every smooth distinct head motion should have the same smooth movement trace. The same smooth movement trace here refers to the track of the movement. We split 78 files for the training data set, 20 files for the validation and 25 files for the test data set.

Training was conducted on a GPU machine and a multi- CPU machine in Tensorflow version 1.12 by mini-batch training using Adam optimisation (learning rate 0.0001)~\cite{Diederik2014}. In the evaluation, the test data set of the same single speaker used in training the prediction model is input to the trained prediction model and the model predicts head motion frame by frame. After that, the output of the prediction model is joined to be distinct head motions and passes them to the proposed filter. The filtered head motion is then processed with the overlap-add method and synthesised. Lastly, the following terms are used for the rest of the discussion in this paper:
\begin{itemize}
    \item \textbf{NonFilter}: Without any filtration
    \item \textbf{ProposedF}: Proposed filter
    \item \textbf{MVA}: Moving average filter
    \item \textbf{GaussianF}: Gaussian filter
\end{itemize}

\section{Objective Evaluation}

We use four metrics to compare the predicted head motion with the ground truth: the mean-squared error (MSE), the local canonical correlation analysis (CCA)~\cite{Haag2016} over 500ms window with 250ms overlap, the absolute SPARC smoothness measure~\cite{Balasubramanian2015} in the speaking region, and the symmetrised Kullback - Leibler (KL) divergence. The first one is a commonly used error metric; the second one is the correlation factor to see how similar in shape the predicted and ground truth graphs are; the third one measures the smoothness of the discrete movements; and the fourth one is used to calculate the relative entropy of the distributions of the movement and evaluate the similarities of the two distributions. Large local CCA represents a high correlation; a small value in the smoothness measure means less movement (zero is the smallest and means no movement at all), and zero in KL divergence indicates that the two distributions are identical. 

In the filter comparison, we would like to ensure that the filtration effect to the movement should be similar while applying different filters in order to ensure that the comparison is fair. As our proposed F is trained and the parameters are fixed, this means that the filtration is same for all input, whereas the linear filters have a scalar parameter to control the impulse response of the filter, to vary how smooth the filter should apply to the input. Thus, this can create different smoothing effects as if the scalar parameter is too small or large for the linear filters. We chose the sigma of the Gaussian filter and the window value of the moving average depending on the ratio of the sum of the power spectrum over 5Hz over the sum of all the power spectrum of the predicted head motion after passing through our proposed filter. As the ratio of these high-frequency motions in the power spectrum over 5Hz is at a similar level after passing the filters, we assume that the noise level is similar. The sigma of the Gaussian filter is then eight, and the window width of the moving average filter is 35.

Lastly, the training data is used to train the model, the validation is used for the parameter optimisation, and the objective evaluations are conducted on the test data set.

\begin{table}[tb]
\centering
\begin{tabular}{|l|c|c|}
\hline
 Model & DN & GN \\
\hline
$150-300-60_{no\_dropout}$ &  \textbf{0.20}&$0.70$\\
$150-300-60_{with\_dropout}$    & $0.22$&\textbf{0.66}\\
\hline
$150-3000-3000_{no\_dropout}$ & \textbf{0.18}&\textbf{0.28}\\
$150-3000-3000_{with\_dropout}$    &  $0.19$&$0.99$\\
\hline
\end{tabular}
\caption{MSE of each model against the noisy datasets. $_{with\_dropout}$ refers to the addition of a dropout layer before the input layer of the model. $_{no\_dropout}$ refers to no dropout layer before the input layer of the model. DN:Dropout Noise, GN:Gaussian Noise.}
\label{tab:architecture_filter}
\vskip -7mm
\end{table}

\subsection{Architecture of the ProposedF}
As the feature dimension is different from the human body motion\cite{GhoshSAH17}, we experimented the same number of layers, but different width of the model to investigate the width of the proposedF and effect of dropping the frames. From Table~\ref{tab:architecture_filter}, we can clearly see that wider models produce better results against the dropout noise under both dropout layer occasions. Moreover, the models without the dropout layer outperform those with the dropout layer in the dropout noise test. On the other hand, model $150-3000-3000_{no\_dropout}$ outperforms other models in the Gaussian noise. 

Next, examining the effect of the smoothing layer, in Table~\ref{tab:middle_filter}, there is no improvement after 180 nodes in the dropout noise and the models that have a wider layer perform worse when testing Gaussian noise. From the above two experiments, we selected $150-3000-180_{no\_dropout}$ architecture to be our proposedF, which is used in the rest experiments.

\begin{table}[tb]
\centering
\begin{tabular}{|c|c|c|c|c|}
\hline
& \multicolumn{2}{c|}{Dropout Noise} & \multicolumn{2}{c|}{Gaussian Noise}\\
\hline
 Node  & Training & Test & Training & Test \\
\hline
60 &  0.30&0.21&\textbf{0.19}&\textbf{0.21}\\
120    & 0.29&0.20&0.20&0.22\\
180 & \textbf{0.28}&\textbf{0.19}&0.21&0.22\\
300    &  0.28&0.19&0.22&0.23\\
3000    &  0.28&0.19&0.27&0.28\\
\hline
\end{tabular}
\caption{MSE of different number of nodes in the smoothing layer for $150-3000-n_{no\_dropout}$ model against the noisy data set created by adding dropout noise (rate=0.5) or Gaussian noise (0, 0.2) to the 'clean' data.}
\label{tab:middle_filter}
\vskip -6mm
\end{table}

\begin{figure}[tb]
\centering
\includegraphics[width=0.3\linewidth]{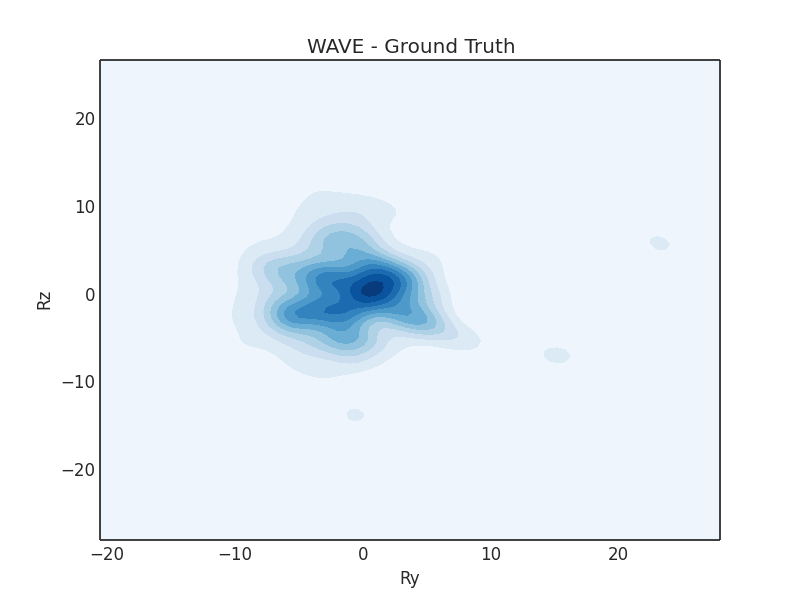}\includegraphics[width=0.3\linewidth]{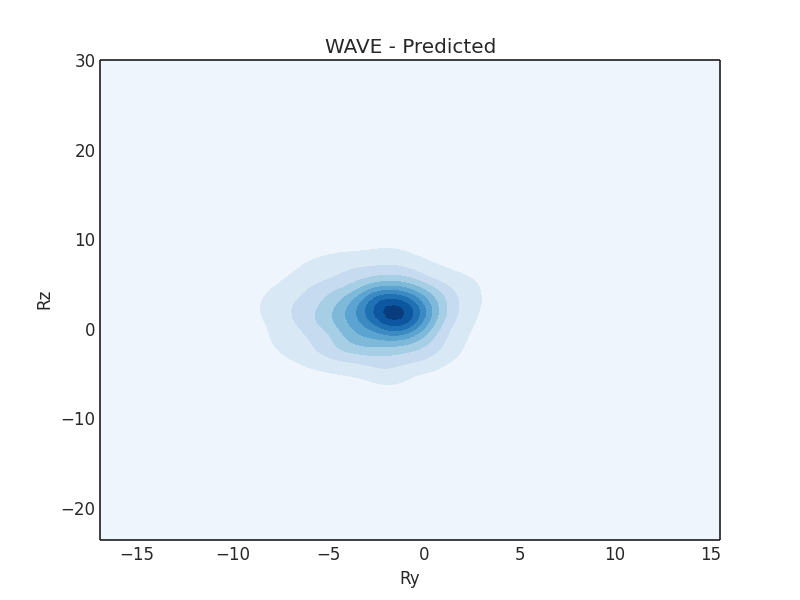}
\includegraphics[width=0.3\linewidth]{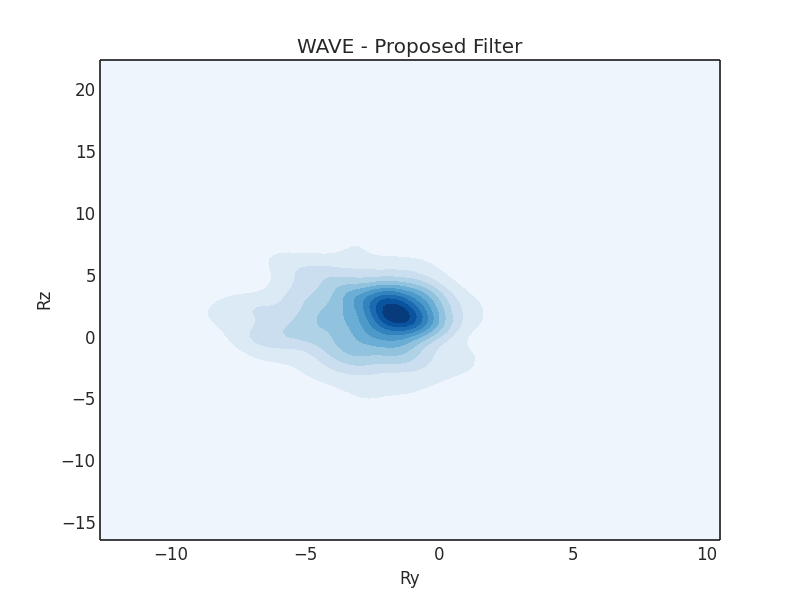}\includegraphics[width=0.3\linewidth]{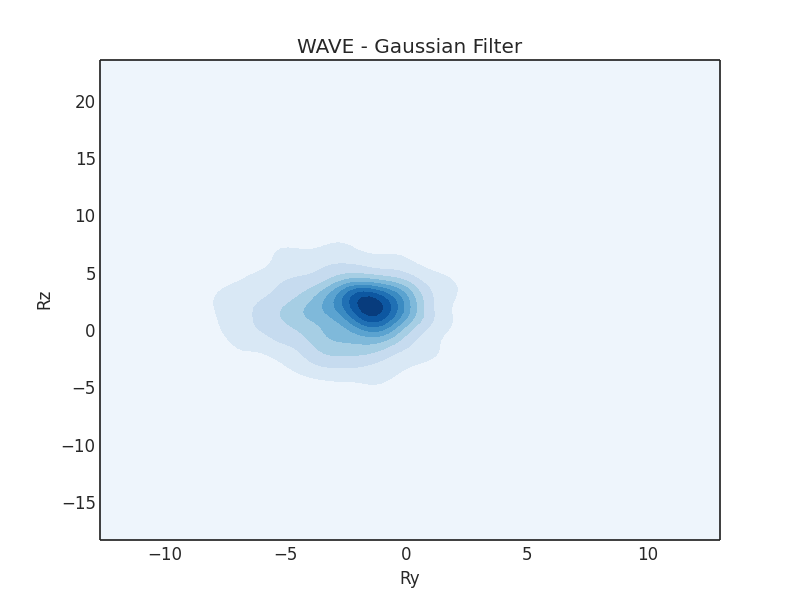}\includegraphics[width=0.3\linewidth]{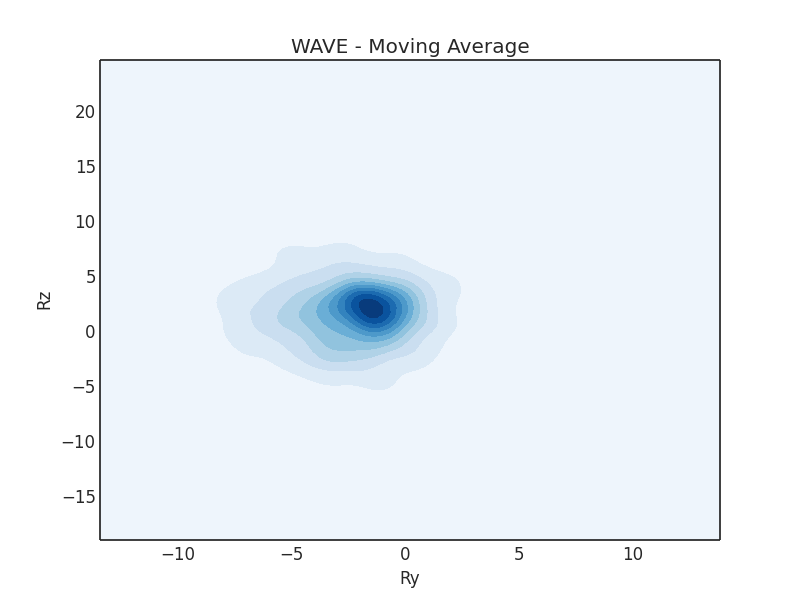}
\caption{The distribution of the Y and Z trajectories in head motion  in the speaking region for the ground truth and the prediction model before(BF)/after(AF) the filters.}
\label{fig:filter_model_distribution}
\vskip -5mm
\end{figure}

\begin{table}[tb]
\vskip -2mm
\centering
\begin{tabular}{|l|c|}
\hline
 Model & KL Divergence\\
\hline
NonFilter & 0.228\\
ProposedF & \textbf{0.168}\\
MVA    & 0.189\\
GaussianF     & 0.175\\
\hline
\end{tabular}
\caption{symmetrised \text{Kullback - Leibler}(KL) divergence between the distribution of the predicted result before/after filtration and the one of the ground truth.}
\label{tab:kl}
\vskip -7mm
\end{table}

\begin{figure}[t]
\centering
\includegraphics[width=0.3\linewidth]{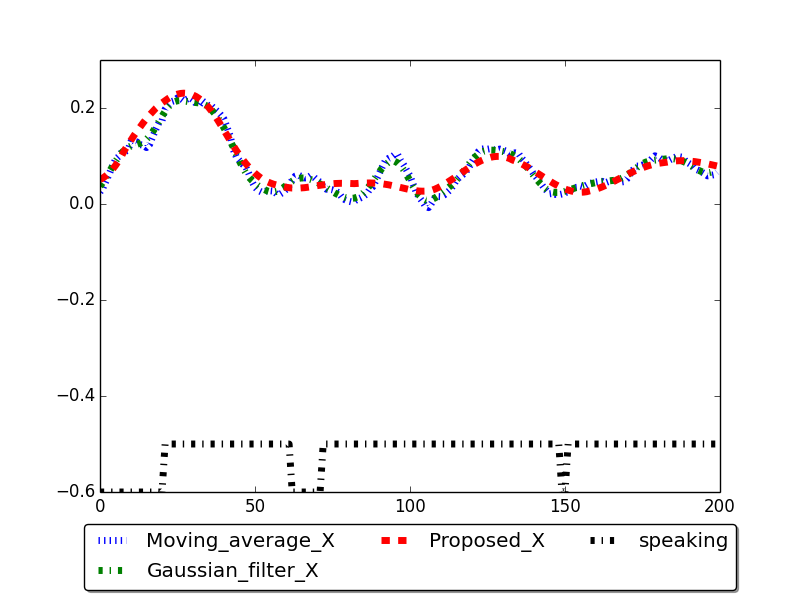}\includegraphics[width=0.3\linewidth]{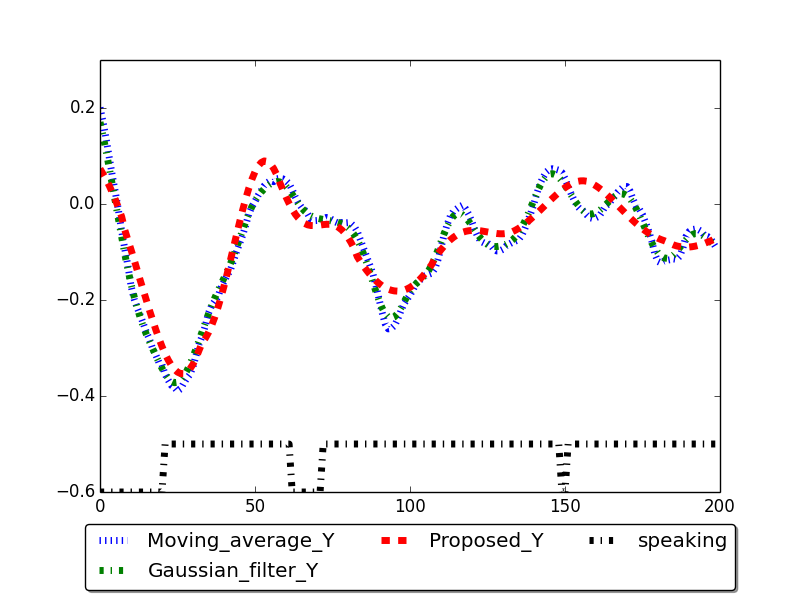}\includegraphics[width=0.3\linewidth]{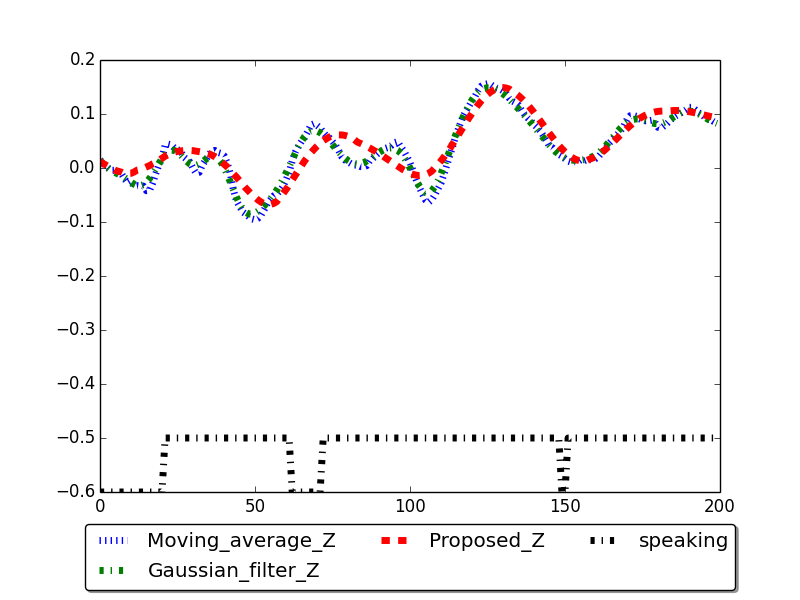}
\caption{The top diagrams show the effect of the different filters de-noising the output from our prediction model in X, Y, Z trajectories. The square-wave plot in each diagram indicates whether the actor is in speaking (up) or listening (down) mode.}
\label{fig:filter_model_effect}
\vskip -1mm
\end{figure}

\subsection{PropoedF VS Linear Filters}
As mentioned in the Sec~\ref{sec:intro}, the linear filter does not have the additional information of the characteristic of the head motion and filters the noise based on the delayed versions of the input signal. We assumed that our proposedF should have learnt the characteristic of the head movement track after training. Therefore, we evaluated our proposed filter and some common linear filters using the result from the prediction model, in order to support our assumption. We can see from Table~\ref{tab:rms_cca}, with the filtration, that the ProposedF achieved the best result as it decreases MSE dramatically from 2.44 to 1.97 and has the highest local CCA of 0.33.

We plotted the distribution of the head motion in Figure~\ref{fig:filter_model_distribution}and observed that the distribution of the ground truth is irregular, whereas the predicted result before filtration is regular, and each filter takes effect to change this. Looking at the same figure of the distribution of the Gaussian filter and moving average method, they have similar shapes and the surrounding of the two centre points are mostly the same, whereas for our proposedF, it tends to change the distribution with some memorised pattern as the overall shape is changed to be irregular in the centre point. Table~\ref{tab:kl} shows that after the filtration, the distribution of the head motion tends to be closer to the ground truth and the one with our proposed filter is the closest with the value of 0.168. 

Figure~\ref{fig:filter_model_effect}, figures on the bottom, illustrates that the predicted result is very noisy and jerky, and it is not close to the ground truth as shown by the sharp edges of the graph. After processing with our proposed filter, the graph is clearly smoother. On the top of the Figure~\ref{fig:filter_model_effect}, we can see that our proposed filter outperforms the other two linear filters as it is less noisy. 

\begin{figure}[t]
\centering
\includegraphics[width=0.3\linewidth]{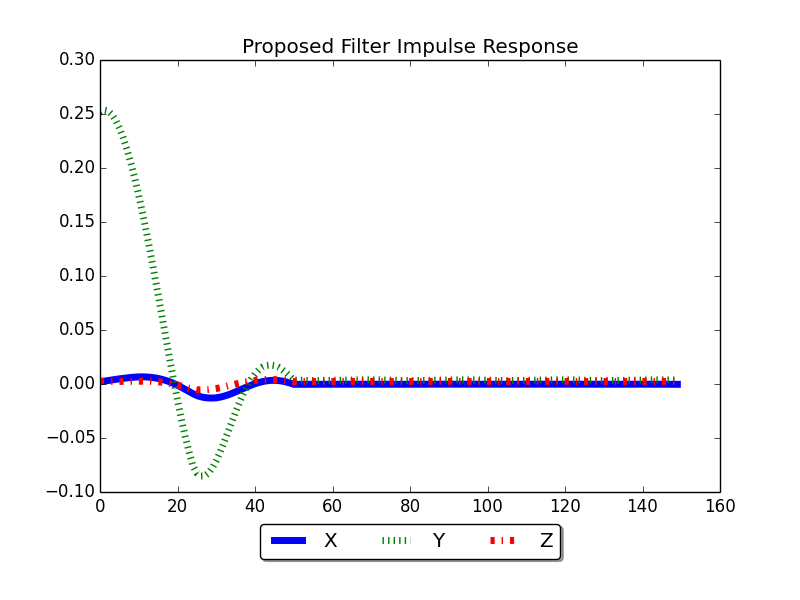}\includegraphics[width=0.3\linewidth]{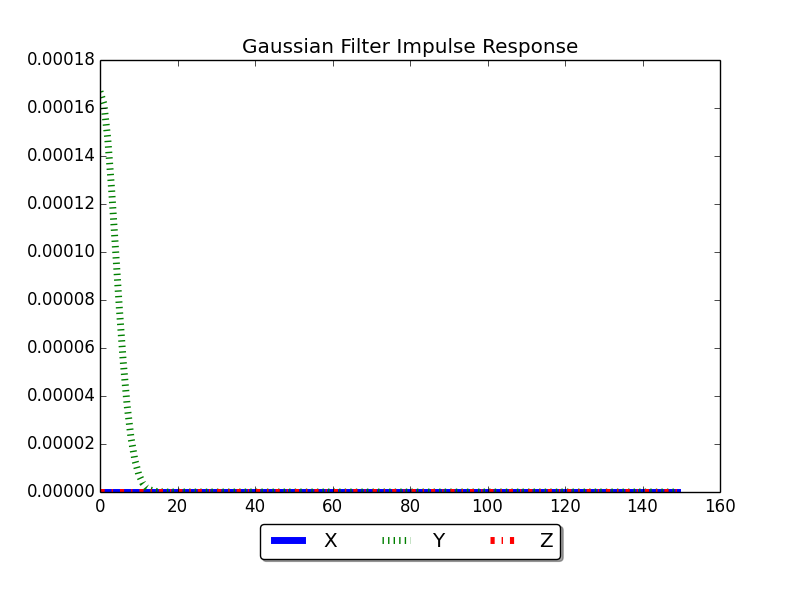}\includegraphics[width=0.3\linewidth]{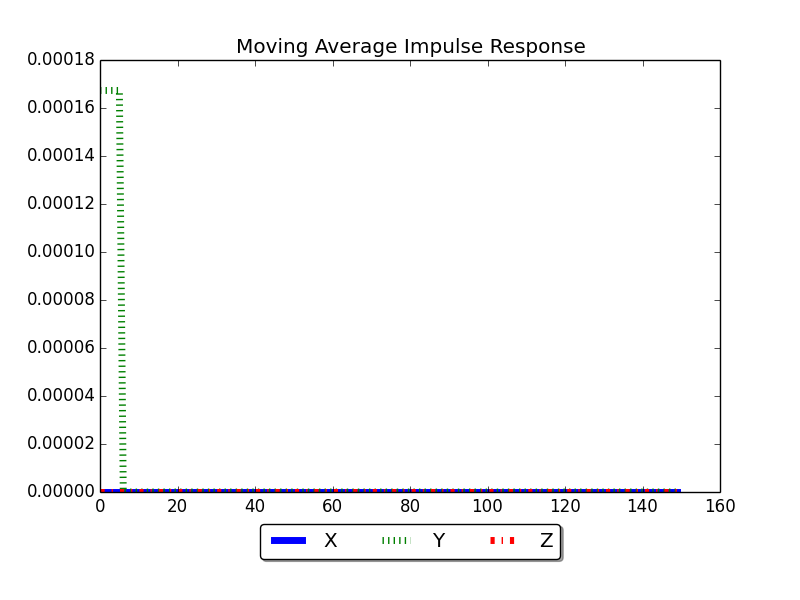}
\caption{Impulse signal on Ry channel effect of each filter. The left one is ProposedF, middle one is the GaussianF and the right one is the MVA.}
\label{fig:filter_model_characteracstic}
\vskip -5mm
\end{figure}

In Figure~\ref{fig:filter_model_characteracstic}, with impulse signal on the Ry channel, linear filters only process that single channel and do not influence the others. However, with our proposed filter, passing in a single channel signal affects two trajectories from time to time. This shows that the filtration of single channel done by proposedF interacts with others two trajectories, and it is the same as the real-life situation where one trajectory of the head moves and the other two trajectories are affected. This proves that our proposedF had learnt the characteristic of the head motion after training. 

\begin{figure}[t]
\centering
\includegraphics[width=\linewidth, height=5cm]{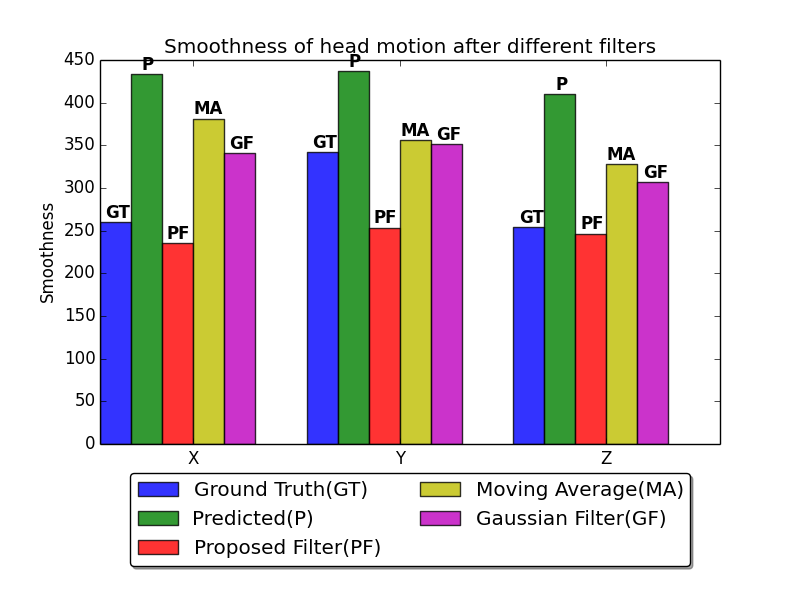}
\caption{Absolute SPARC smoothness value for head motion rotation vector (X, Y, Z) in the speaking region before/after the filters. Higher smoothness value refers to there being more movements in the same time period.}
\label{fig:smoothness}
\vskip -8mm
\end{figure}

Figure~\ref{fig:smoothness} shows the absolute SPARC smoothness values decrease after the filter. It is clear that our proposed filter has stronger filtration effect than the other two linear filters as the absolute smoothness values in three trajectories decrease the most from the predicted result.

\begin{table}[th]
\centering
\begin{tabular}{|l|c|c|c|c|}
\hline
&\multicolumn{2}{c|}{Before filter}&\multicolumn{2}{c|}{After filter} \\
\hline
 Model & MSE & CCA & MSE & CCA \\
\hline
ProposedF && & \textbf{1.97}&\textbf{0.33}\\
MVA    & $2.44$& $0.28$& $2.20$&$0.32$\\
GaussianF    & & & $2.15$&$0.33$\\
\hline
\end{tabular}
\caption{ MSE and local CCA in speaking region for each model before/after the de-noising filter.}
\label{tab:rms_cca}
\vskip -8mm
\end{table}

\begin{figure}[tb]
\centering
\includegraphics[width=\linewidth, height=7cm]{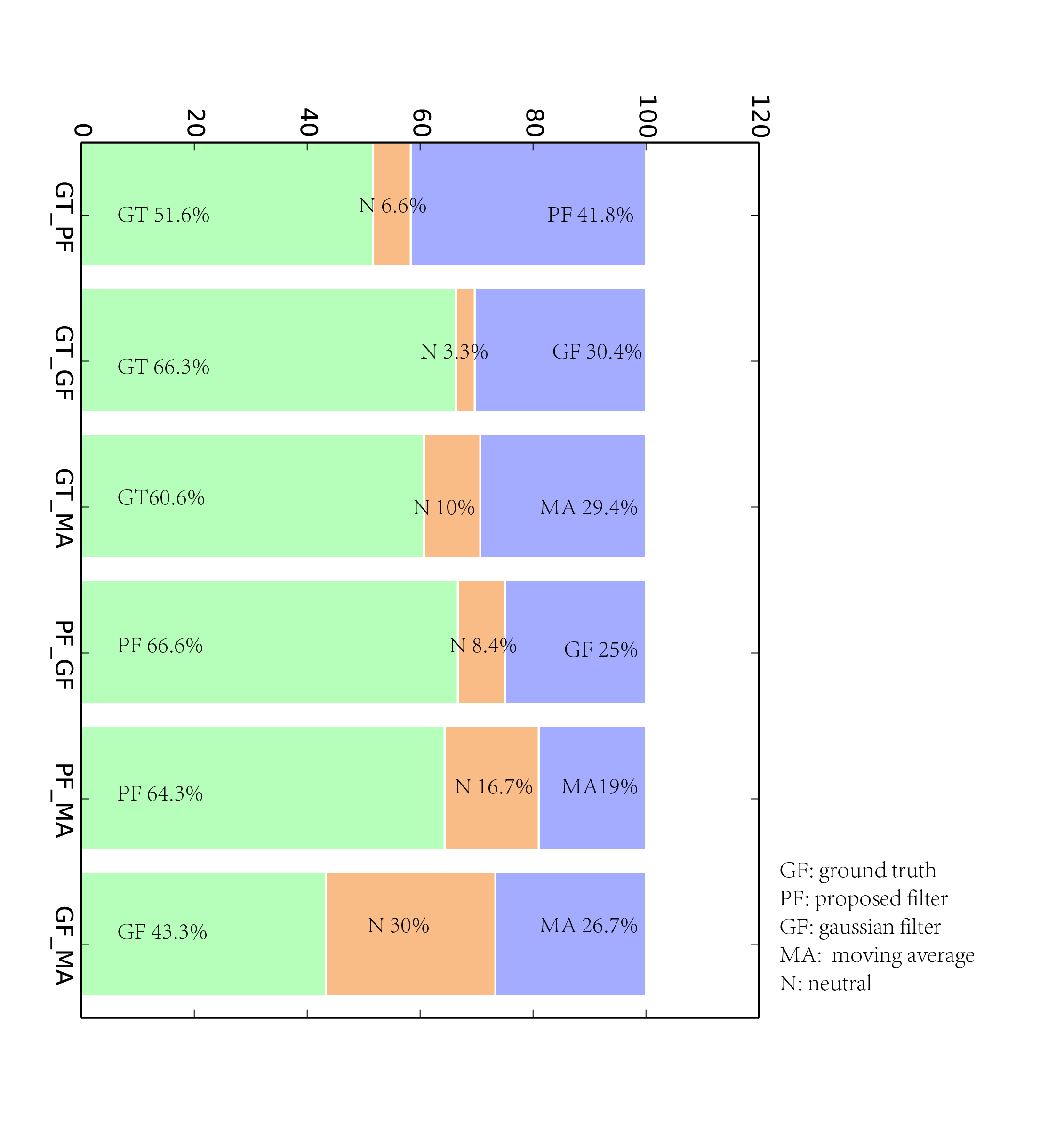}
\caption{The percentage preference of A/B test.}
\label{fig:subjective}
\vskip -6mm
\end{figure}

\section{Subjective Evaluation}
We conducted A/B preference tests on the naturalness of the synthesised head motion animation. We randomly selected 15 speaking regions in an audio file and split them into five test groups. Each test group has a total of 18 comparison tests comparing between ground truth, head motion filtered by Pro- posedF, head motion filtered by GaussianF, and head motion filtered by MVA. A group of 20 participants were involved in this evaluation, and they were asked to decide which one was better according to the naturalness of the head motion. 

The evaluation result is presented in Figure~\ref{fig:subjective}. We can see that participants were unable to tell the difference between the ground truth and the proposed filter, but they could easily pick that ground truth is better as compared to the Gaussian filter or moving average. While comparing proposedF and GaussianF (or MVA), participants preferred to choose the proposedF. Lastly, GaussianF is slightly preferred over moving average. However, the neutral is the highest among the six comparison tests, indicating that participants thought both of them are highly similar.

\section{Conclusions}

In this paper, we have studied an effective head motion filter by reconstructing the head movement track in the training stage. We described our data, evaluated the feasibility of our filter model, and compared the filtration effect with common linear filters. From extensive evaluations, we can conclude that (1) an appropriate number in the middle layer of the filter model is essential to reconstruct head motions against the noise. (2) Our proposed filter demonstrates good smoothing effect to the noisy data and the predicted head motion with large error dropping. (3) Objective evaluation results show that our proposed filter has the capability to recover the motion movement, taking advantage of the filtration process by knowing the characteristic head motion, as compared to the common linear filters which do not do so. (4) Subjective evaluation reveals that participants preferred to choose the head motion processed with our proposedF over the other two filters. In the future, we would like to investigate a better prediction model in the head motion synthesis by applying our proposed filter to a neural network.


\end{document}